\documentclass[sigconf]{acmart}
\usepackage{comment}
\usepackage{xcolor}
\usepackage[most]{tcolorbox}
\usepackage{hyperref}
\usepackage{xspace}
\usepackage[scaled=0.85]{beramono}
\usepackage{bm}
\usepackage{tikz}
\usetikzlibrary{arrows.meta, positioning, shapes.geometric,automata}
\tikzset{>=Stealth}

\usepackage{listings}
\usepackage{lstautogobble} 
\usepackage{csquotes}
\usepackage{siunitx}
\usepackage{cleveref} 
\usepackage{subcaption}
\usepackage[inline]{enumitem}

\newtcolorbox{graybox}{
  colback=lightgray,
  colframe=black,
  boxrule=0.5pt,
  arc=2pt,
  left=5pt,
  right=5pt,
}

\newcommand{\numscenarios}{$7$\xspace}

\newenvironment{researchquestion}[2]{%
  \par\noindent
  \textbf{(#1)\label{rq:#1} #2. }\ignorespaces
}{\par}

\newcommand{\rqref}[1]{\hyperref[rq:#1]{#1}}

\newcommand{\roadlogic}{\textsc{RoadLogic}\xspace}
\newcommand{\clingo}{\textsc{clingo}\xspace}

\newcommand{\Actors}{A}
\newcommand{\actor}{a}

\newcommand{\pos}{x}
\newcommand{\vel}{v}

\newcommand{\road}{r}

\newcommand{\lane}{l}

\newcommand{\letter}{\sigma}
\newcommand{\trace}{\tau}

\newcommand{\drive}{\textsf{drive}}

\newcommand{\at}{@}
\newcommand{\tbegin}{\textsf{begin}}
\newcommand{\tend}{\textsf{end}}

\newcommand{\relation}{P}

\newcommand{\sa}{\mathcal{A}}

\newcommand{\algebra}{\mathbb{A}}
\newcommand{\sem}[1][]{\mbox{$[\!\bf[ #1 ]\!]$}}
\newcommand{\domain}{\mathcal{D}}
\newcommand{\Predicates}{\mathcal{P}}

\newcommand{\carla}{CARLA\xspace}
\newcommand{\openscenario}{OpenSCENARIO\xspace}

\lstdefinelanguage{OS2}{
    keywords={scenario, do},
    keywords=[2]{parallel, drive, serial, lane, position},
    keywordstyle={\color{blue}\bfseries},
    keywordstyle=[2]{\color{red!50!blue}},
    sensitive=false, 
    morecomment=[l]{\#}, 
    showstringspaces=false,
    numbers=left,
    stepnumber=1,
} %

\newcommand{\ego}{\texttt{v1}\xspace}
\newcommand{\car}{\texttt{v2}\xspace}

\newcommand{\osTwo}{OS2\xspace}

\newcommand{\AutT}[1]{\mathcal{T}(#1)} 
\newcommand{\parcomp}{\mathbin{\parallel}} 

\newcommand{\kwdriveshort}[2]{\texttt{drive}(\texttt{start}:#1,\, \texttt{end}:#2)\xspace}

\lstdefinelanguage{ASP}{
  morekeywords={not,\#show,\#const,\#program,\#external},
  sensitive=true,
  morecomment=[l]{\%},
  morestring=[b]",
  alsoletter={\#}
}

\lstdefinestyle{asp}{
  language=ASP,
  basicstyle=\ttfamily\footnotesize,
  columns=fullflexible,
  breaklines=true,
  breakatwhitespace=false,
  keepspaces=true,
  showstringspaces=false,
  autogobble=true
}

\newtcblisting{aspbox}[1][]{%
  listing only,
  listing engine=listings,
  listing options={
    style=asp, 
    escapechar=@,
  },
  colback=black!2,
  colframe=black!40,
  boxsep=1mm,
  top=1mm, bottom=1mm, left=1mm, right=1mm,
  enhanced,
  sharp corners,
  #1
}

\title{Declarative Scenario-based Testing with \roadlogic}

\author{Ezio Bartocci}
\email{ezio.bartocci@tuwien.ac.at}
\affiliation{%
  \institution{TU Wien}
  \city{Vienna}
  \country{Austria}
}

\author{Alessio Gambi}
\email{alessio.gambi@ait.ac.at}
\affiliation{%
  \institution{AIT Austrian Institute of Technology}
  \city{Vienna}
  \country{Austria}
}

\author{Felix Gigler}
\email{felix.gigler@ait.ac.at}
\affiliation{%
  \institution{TU Wien}
  \city{}
  \country{}
}
\affiliation{%
  \institution{AIT Austrian Institute of Technology}
  \city{Vienna}
  \country{Austria}
}

\author{Cristinel Mateis}
\email{cristinel.mateis@ait.ac.at}
\affiliation{%
  \institution{AIT Austrian Institute of Technology}
  \city{Vienna}
  \country{Austria}
}

\author{Dejan Ničković}
\email{dejan.nickovic@ait.ac.at}
\affiliation{%
  \institution{AIT Austrian Institute of Technology}
  \city{Vienna}
  \country{Austria}
}


\setcopyright{none}
\renewcommand\footnotetextcopyrightpermission[1]{}

\acmConference[]{}{}{}
\acmBooktitle{}
\acmDOI{}
\acmISBN{}

\makeatletter
\fancypagestyle{plain}{%
  \fancyhf{}%
  \fancyfoot[C]{\thepage}%
}
\makeatother

\begin{document}

\begin{abstract}
Scenario-based testing is a key method for cost-effective and safe validation of autonomous vehicles (AVs).
Existing approaches rely on imperative scenario definitions, requiring developers to manually enumerate numerous variants to achieve coverage.
Declarative languages, such as ASAM \openscenario~DSL (OS2), raise the abstraction level but lack systematic methods for instantiating concrete and specification-compliant scenarios. 
To our knowledge, currently, no open-source solution provides this capability.

We present \roadlogic that bridges declarative OS2 specifications and executable scenarios. It uses Answer Set Programming to generate abstract plans satisfying scenario constraints, motion planning to refine the plans into feasible trajectories, and specification-based monitoring to verify correctness.

We evaluate \roadlogic on instantiating representative OS2 scenarios executed in the CommonRoad framework.
Results show that \roadlogic consistently produces realistic, specification-satisfying simulations within minutes and captures diverse behavioral variants through parameter sampling, thus opening the door to systematic 
scenario-based testing for autonomous driving systems.

\end{abstract}

\maketitle
\pagestyle{plain}
\thispagestyle{plain}

\section{Introduction}

Testing autonomous vehicles (AVs) is inherently complex and requires validating AVs in a broad range of driving scenarios, 
including those involving other road users. Conducting exhaustive testing in the real world is costly, impractical, and dangerous~\cite{kalra2016driving}. Therefore, \emph{simulation-based testing} has become an essential approach for scalable, systematic AV validation, enabling engineers to safely and efficiently evaluate critical and rare situations.

Thorough testing of AVs in simulation requires the ability to systematically describe, reproduce, and analyze a wide range of potentially safety-critical driving situations. To address this challenge, \emph{scenario-based testing} has emerged as a key methodology for validating AVs, offering a powerful means to define test cases in a rigorous yet human-interpretable way. Originally introduced in the PEGASUS project~\cite{pegasus}, scenarios come in four levels of abstraction: functional, abstract, logical, and concrete scenarios \cite{menzel2018scenarios}. 
Functional scenarios are behavior-based, structured descriptions of traffic situations in natural language. 
Abstract scenarios formalize functional scenarios by describing them declaratively. 
Logical scenarios refine abstract scenarios as a parameterized set of traffic situations.
Concrete scenarios are executable instances of a logical scenario, with defined scenery, parameters, and road users' behavior. 

In recent years, considerable effort has been devoted to instantiating scenario-based testing for AVs using scenario description languages and integrating these languages with driving simulators.
Prominent examples include ASAM~\openscenario~XML~\cite{asam_openscenario_userguide_v1_0}, a concrete and executable scenario language, and Scenic~\cite{fremont2019scenic}, a probabilistic scenario language, both integrated with \carla~\cite{Dosovitskiy17}.
However, these scenario languages are operational and require detailed specification of how individual scenarios are executed rather than supporting higher-level, more abstract representations of driving situations. 

\begin{lstlisting}[
    float=tp,
    floatplacement=tbp,
    basicstyle=\ttfamily\normalsize,
    language=OS2,
    caption={An overtake scenario in OpenSCENARIO DSL.},
    label={lst:os2}
]
scenario traffic.overtake:
  v1: car
  v2: car

  do parallel():
    v2.drive()
    serial:
      A: v1.drive() with:
        lane(same_as: v2, at: start)
        lane(left_of: v2, at: end)
        position([10..20]m, behind: v2, at: start)
      B: v1.drive() with:
        position([1..10]m, ahead_of: v2, at: end)
      C: v1.drive() with:
        lane(same_as: v2, at:end)
        position([5..10]m, ahead_of: v2, at: end)
\end{lstlisting}
\begin{figure}
\centering
\includegraphics[width=\columnwidth]{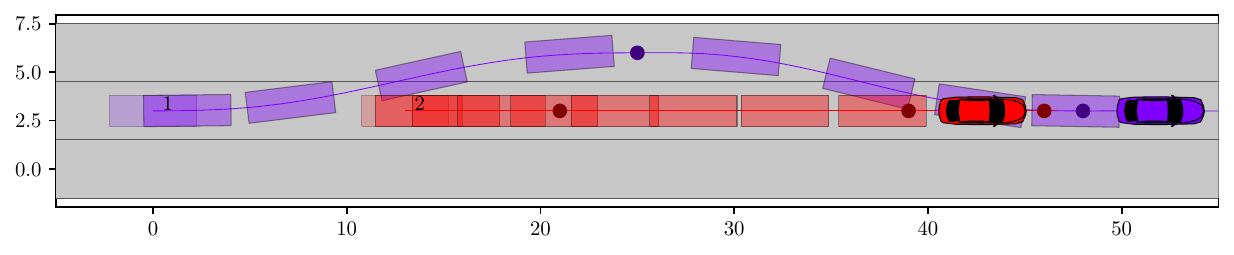}
\caption{Visualization of a simulation compliant with the specification in Listing~\ref{lst:os2} generated by \roadlogic. \emph{v1} is the (purple) vehicle performing the overtake, while \emph{v2} is the (red) vehicle driving straight.}
\Description{Overtaking scenario with two vehicles on adjacent lanes.}
\label{fig:working-example}
\end{figure}

The ability to specify high-level dynamic traffic scenarios is addressed by ASAM~\openscenario~DSL (OS2)~\cite{os2.1_spec},
a declarative language for describing abstract scenarios. OS2 defines the behavior of all traffic participants, including the ego vehicle and non-playable characters (NPCs). 
For instance, Listing~\ref{lst:os2} illustrates a specification of an overtake scenario in OS2 in which \emph{v1} initially drives behind \emph{v2} on the same lane (lines~$8-11$) and finally drives in front of it (lines~$14-16$). 
By separating the \emph{what} (the intended situation) from the \emph{how} (its execution in simulation), OS2 enables concise expression of abstract scenarios that possibly encompass many valid concrete scenarios, thus promoting automation, reuse, and traceability across the AV testing pipeline. However, 
OS2 lacks systematic open-source methods for instantiating specification-compliant simulations, such as the one reported in Figure~\ref{fig:working-example}.

This paper introduces \roadlogic, the first open-source framework that automatically generates 
simulations from OS2 scenario descriptions. The proposed approach consists of the following steps: (1) parsing the input OS2 scenario description into a symbolic automaton representation, (2) encoding the symbolic automaton into an Answer Set Programming (ASP) problem, (3) resolving the ASP problem using an off-the-shelf solver to produce a high-level plan, (4) concretizing the high-level plan into a realistic simulation using an AV planner, and (5) ensuring that the concrete scenario is compliant with the abstract scenario specification, using a runtime monitor generated from its symbolic automaton representation. 

Unlike operational scenario languages such as Scenic~\cite{fremont2019scenic}, \roadlogic targets OS2 specifications and relies on explicit symbolic reasoning to derive concrete simulations, rather than on probabilistic generation alone. In contrast to proprietary OS2 toolchains, \roadlogic is fully open-source and exposes the complete generation process, from symbolic planning to simulation and monitoring.

We implemented the \roadlogic framework on top of the \clingo ASP solver~\cite{DBLP:journals/corr/GebserKKS14}, the FrenetiX motion planner~\cite{Raineretal2024Frenetix}, and the CommonRoad simulation environment~\cite{commonroad2017}, and we evaluated it across a variety of logical scenarios.
As our evaluation shows, \roadlogic effectively bridges the gap between high-level declarative scenario descriptions and concrete autonomous vehicle simulations.

This paper makes the following main contributions:
\begin{itemize}
    \item We introduce \roadlogic, an open-source framework that automatically instantiates declarative OS2 specifications into realistic and specification-compliant simulations by leveraging ASP and AV motion planning.
    \item We deliver an end-to-end pipeline where the OS2 specifications drive both concrete scenario generation and formal conformance monitoring, ensuring that only simulations faithful to the abstract scenario specification are retained.
    \item We evaluate \roadlogic on diverse OS2 scenarios, showing that it efficiently produces realistic and specification-compliant simulations bridging declarative scenario design and concrete AV testing.
\end{itemize}

In the remainder of the paper, 
Section~\ref{sec:background} introduces the necessary background;
Section~\ref{sec:method} presents the main contributions of this work;
Section~\ref{sec:evaluation} details the evaluation of our approach, followed by Section~\ref{sec:related}, which discusses related work.
Section~\ref{sec:discussion} critically discusses design choices and implications of our 
framework.
Finally, Section~\ref{sec:conclusions} concludes the paper.

\section{Background}
\label{sec:background}

\subsection{OpenSCENARIO DSL}

OpenSCENARIO DSL is a high-level, human-interpretable domain-specific language for declaratively specifying driving scenarios.
This section limits the description to the fragments of the language that we use in this paper. 

An OS2 specification defines a system of \emph{actor} objects that have a set of fields defined by their type definition.
Let $\Actors = \{ \actor_1, \ldots, \actor_n \}$ denote a set of actors. We denote by $\road(m)$ a road with $m$ lanes, where $0$ is the leftmost and $m-1$ the rightmost lane. Given an actor $\actor_i \in \Actors$, we denote by $\pos(\actor_i)$ its position, $\vel(\actor_i)$ its velocity and $\lane(\actor_i)$ the lane of the actor $\actor_i$ on the road $\road(m)$.

A state $s$, defined over the set of actors $A$ and the road $\road(m)$, is the 
tuple $(\pos(\actor_1), \vel(\actor_1), \lane(\actor_1), \ldots, \pos(\actor_n), \vel(\actor_n), \lane(\actor_n), t)$ defining the position, velocity and lane occupancy of every actor $\actor_i \in \Actors$ on the road $\road(m)$ at time $t$. 
A trace $\trace~:~\mathbb{R}_{\geq 0} \to S$ is a continuous time sequence of states, where $S$ is a set of states. We assume that in practice, a trace $\tau$ is given as a discrete sequence $s_1, \ldots, s_n$ of (time-stamped) states with constant-interpolation between the sampled states.
Given a set of actors $\Actors$, a road $\road(m)$, we define the syntax of the fragment of OS2\footnote{We use the mathematical notation for the syntax definition instead of the actual syntax used in Listing~\ref{lst:os2} to optimize space.} that we use in the paper as follows:

$$
\begin{array}{ll}
\varphi := & \phi ~|~ \varphi_1~||~\varphi_2~|~ \varphi_1~\cup~\varphi_2 
\\

\phi := & \psi~|~\phi_1 \cdot \phi_2 
\\
\psi := & \drive(\actor)~|~\drive(\actor)~:~\gamma
\\
\gamma := & (\relation(S), \at)~|~\gamma_1 \wedge \gamma_2 \\
\end{array}
$$
\noindent where $\relation$ is an $n$-ary predicate over actor states $S$, 
$\at \in \{ \tbegin, \tend \}$. Intuitively, this fragment of OS2 consists of a sequence of driving behaviors ($\drive(\actor)$) that can be restricted with \emph{constraints} ($\gamma$) over (pairs of) state variables defined at the beginning and the end of each segment. The atomic behaviors can be combined into more complex scenarios using \emph{serial} ($\cdot$), \emph{parallel} ($||$) and \emph{one-of} ($\cup$) composition. 

An OS2 scenario $\varphi$ is evaluated over a finite trace $\trace$ defined over an interval $[t, t')$. 
We use $\tau[t]$ to denote the state of the system at time $t$ in trace $\tau$, and $x(\tau[t])(a)$ denotes the longitudinal position of actor $a$ in that state.
The semantics of OS2 is defined inductively as follows:
$$
\begin{array}{lcl}
(\tau, t, t') \models \varphi_1~||~\varphi_2 & \leftrightarrow & (\tau, t, t') \models \varphi_1 \textrm{ and } (\tau, t, t') \models \varphi_2 \\

(\tau, t, t') \models \phi_1 \cdot \phi_2 & \leftrightarrow & \exists t'' \in (t, t')  \textrm{ s.t. } (\tau, t, t'') \models \phi_1 \\
&& \textrm{ and }  (\tau, t'', t') \models \phi_2 \\

(\tau, t, t') \models \varphi_1~\cup~\varphi_2 & \leftrightarrow & (\tau, t, t') \models \varphi_1 \textrm{ or } (\tau, t, t') \models \varphi_2 \\

(\tau, t, t') \models \drive(\actor)~:~\gamma & \leftrightarrow & \forall t'' \in (t+1, t')  x(\tau[t'']) \\
&& (\actor) > x(\tau[t'' - 1])(\actor) \textrm{ and }  \\
&& x(\tau, t, t') \models \gamma\\
(\tau, t, t') \models (\relation(S), \tbegin) & \leftrightarrow & \tau[t] \models \relation(S) \\
(\tau, t, t') \models (\relation(S), \tend) & \leftrightarrow & \tau[t'] \models \relation(S) \\
\end{array}
$$

The above rules define the semantics of OS2 scenarios in terms of trace satisfaction. 
Parallel composition ($\varphi_1 \parallel \varphi_2$) requires both scenarios to hold over the same interval, while 
sequential composition ($\varphi_1 \cdot \varphi_2$) splits the interval into two consecutive segments where each subscenario holds.
The choice operator ($\varphi_1 \cup \varphi_2$) expresses alternative behaviors by requiring that at least one of the two scenarios is satisfied.
The rule for $\textit{drive}(a):\gamma$ captures the intended meaning of a driving behavior: the actor $a$ must make continuous forward progress along the road (expressed by the monotonic increase of its longitudinal position), while the associated constraints $\gamma$ must hold at the specified temporal anchor (either at the beginning or the end of the execution segment).
The last two rules define how predicates are evaluated on the trace at these temporal anchors.

\subsection{Symbolic Automata}

A symbolic finite automaton (SFA)~\cite{DBLP:conf/icst/VeanesHT10} is a state machine similar to a regular finite-state automaton. However, transitions in an SFA are not labeled with alphabet letters, but rather with predicates defined over variables that describe whole (possibly infinite) sets of inputs at once. We formalize this idea with an \emph{effective Boolean algebra} $\algebra$, a tuple 
$(\domain, \Predicates, \sem[\cdot], \top, \bot, \vee, \wedge, \neg)$, where 
$\domain$ is a set of {\em domain elements},  $\Predicates$ is a set of predicates closed under Boolean operations, 
$\sem[\cdot]~:~\Psi \to 2^{\domain}$ is 
a {\em denotation function} such that for all $P, P_{1}, P_{2} \in \Predicates$, 
$\sem[\bot] = \emptyset$, $\sem[\top] = \domain$, 
$\sem[P_{1} \vee P_{2}] = \sem[P_1]\cup \sem[P_{2}]$, 
$\sem[P_{1} \wedge P_{2}] = \sem[P_1]\cap \sem[P_{2}]$, and 
$\sem[\neg P] = \domain \backslash \sem[P]$.

A symbolic finite automaton $\sa$ is then a tuple
$(\algebra, Q, q_{0}, F, \Delta)$, where $\algebra$ is an effective Boolean algebra, 
$Q$ is a finite set of {\em locations}, $q_{0} \in Q$ is an {\em initial} location, 
$F\subseteq Q$ is a set of {\em final} locations, and $\Delta \subseteq Q \times \Predicates \times Q$ 
is the {\em transition} relation.

An SFA starts in an initial location and reads an input one piece at a time; at each step, it checks which transition’s condition is satisfied by the current input element, follows that transition, and moves to a new location. If, after consuming the entire input, the automaton ends up in one of its accepting locations, the input is considered accepted; otherwise, it is rejected. This way, the SFA behaves like a traditional automaton but uses logical conditions to express transitions more compactly.

A {\em finite word} $w$ over 
$\domain$ is a finite sequence $\letter_{1}\letter_{2} \ldots \letter_{n}$ of letters in $\domain$. We denote by $|w|$ the 
length $n$ of the word $w$. Given a letter $\letter \in \domain$, a $\letter$-transition from $q$ to $q'$, denoted by 
$q \xrightarrow{\letter} q'$, is a transition $(q, \psi, q') \in \Delta$ such that $\letter \in \sem[P]$. 
We say that a SFA $A$ {\em accepts} a word $w = \letter_{1}\letter_{2}\ldots \letter_{n}$ if there exists a sequence of 
$\letter$-transitions
$$
q_{0} \xrightarrow{\letter_{1}} q_{1} \xrightarrow{\letter_{2}} q_{2} \xrightarrow{\letter_{3}} \cdots \xrightarrow{\letter_{n}} q_{n}, 
$$
\noindent such that $q_{i} \in Q$, $q_{0}$ is the initial location and $q_{n} \in F$. The {\em language} $L(\sa)$ of 
$\sa$ is the set of all words that $A$ accepts.

We use symbolic automata as an intermediate language for representing OS2 scenarios. The symbolic automata encoding of the scenarios play a twofold role: to derive the high-level plans and to monitor the compliance of the concrete simulations to the abstract scenario intent. 

\subsection{Answer Set Programming}
\label{sec:asp}
    
\textit{Answer Set Programming} (ASP) is a declarative problem-solving paradigm rooted in non-monotonic logic programming and the stable model semantics~\cite{DBLP:conf/iclp/GelfondL88, DBLP:journals/cacm/BrewkaET11, DBLP:books/sp/Lifschitz19}.
An ASP program consists of a finite set of logical rules of the general form
\begin{equation}
a \leftarrow b_1, \ldots, b_m, \operatorname{not}\ b_{m+1}, \ldots, \operatorname{not}\ b_n
\label{eq:asp-rule}
\end{equation}
where \(a\) and \(b_i\) are atoms that may contain variables. 
In practice, an ASP program consists of a collection of \emph{facts} and \emph{rules}. Facts are ground atoms that encode the concrete problem instance (e.g., initial states, vehicles, or road structure), while rules such as Eq.~\eqref{eq:asp-rule} specify logical constraints and state transitions. Together they form the input given to the ASP solver. Eq.~\eqref{eq:asp-rule} illustrates the general rule schema; the concrete instance of the problem is provided separately through ground facts elsewhere in the program.
Before reasoning, an ASP solver performs a \emph{grounding} step that replaces all variables with constants appearing in the program, producing a variable-free (ground) program.
Each \emph{answer set} (or \emph{stable model}) of the grounded program represents a consistent collection of atoms that jointly satisfy all rules.
The symbol ``\texttt{not}'' denotes \emph{default negation} (negation as failure), meaning that a literal is assumed false unless it can be proven true within the current model.

ASP naturally supports combinatorial search, constraint reasoning, and discrete-time planning~\cite{DBLP:series/synthesis/2012Gebser}.
Beyond standard rules, ASP provides several expressive constructs:
\emph{choice rules} allow the non-deterministic generation of candidate actions or states,
\emph{integrity constraints} (rules with empty heads) eliminate invalid combinations, and
\emph{weak constraints} enable expressing optimization preferences such as minimizing trajectory length or deviation.
Modern solvers such as \clingo~\cite{DBLP:journals/corr/GebserKKS14} and \textsc{DLV}~\cite{leone2006dlv} combine expressive modeling languages with advanced conflict-driven learning and optimization techniques derived from modern SAT and constraint solving.
\clingo further supports \emph{multi-shot solving}~\cite{DBLP:journals/tplp/GebserKKS19}, which incrementally extends existing solution candidates with additional steps or constraints without recomputation from scratch.
These features, especially incremental reasoning, make ASP well-suited for generating driving maneuvers under declarative scenario specifications.

In this work, ASP serves as the symbolic reasoning engine for instantiating abstract driving scenarios.
Given a symbolic automaton derived from an OS2 specification, we generate an ASP instance encoding the specification constraints.
Combined with a predefined background theory that captures common driving rules and simplified dynamics, the multi-shot solver computes answer sets representing feasible discrete plans that satisfy the declarative scenario constraints.
These plans form the symbolic basis for subsequent trajectory refinement and physically realistic simulation.

It is worth briefly contrasting ASP with Satisfiability Modulo Theories (SMT), another commonly used paradigm for constraint solving. SMT solvers are particularly effective for reasoning over rich numeric theories such as linear arithmetic or bit-vectors, but typically assume a fixed problem horizon and require all constraints to be encoded upfront. In contrast, ASP is well suited for combinatorial search and non-monotonic reasoning, and naturally supports incremental discrete planning through mechanisms such as multi-shot solving. These properties make ASP particularly convenient for expressing evolving maneuver sequences and exploring alternative scenario realizations under declarative constraints.

\subsection{Action Languages and Planning}
\label{sub:actionlanguages}

Action languages provide a formal basis for reasoning about actions and change.
Originally introduced by Gelfond and Lifschitz~\cite{gelfond1998action}, they serve as \enquote{formal models of parts of natural language that are used for talking about the effects of actions.}
In essence, an action language describes how the execution of actions transforms a system from one state to another.

Formally, a transition system consists of a set of fluents (propositional variables describing properties of the world), a set of actions, and a transition relation connecting states before and after an action is executed.
A \emph{planning problem} then asks for a sequence of actions that leads from an initial state to a goal state, given the system's dynamics.

Languages such as $\mathcal{A}$ and $\mathcal{B}$~\cite{gelfond1998action} define rules of the form \enquote{action $\alpha$ causes fluent $f$ if condition $\varphi$ holds,} which can be directly represented in ASP using predicates such as \texttt{holds(f, t)} and \texttt{occurs(alpha, t)}.
This encoding allows us to 
leverage ASP solvers for planning, a paradigm 
known as \emph{answer set planning} (see~\cite{son2006domain, actionsurvey_2023}). 

In practice, this integration allows declarative planning problems to be expressed compactly: initial states are represented as facts, dynamic causal laws as transition rules, and goal conditions as constraints.
Solvers such as \clingo then compute sequences of actions, i.e., \emph{plans}, that satisfy all specified preconditions and effects.

We follow this tradition to express driving dynamics and scenario evolution in our framework, using an ASP encoding that captures both discrete vehicle actions and their causal effects on system state.

\subsection{CommonRoad and FrenetiX}
\label{sec:commonroad}

\roadlogic instantiates scenarios in CommonRoad~\cite{commonroad2017} and execute them against the FrenetiX motion planner~\cite{Raineretal2024Frenetix} using Kaufeld et al.'s multi-agent platform~\cite{multiagent2024} to ensure scenarios realism. 

CommonRoad 
is an established framework for motion planning development, which is central to autonomous driving.
CommonRoad scenarios comprise three main elements:
(i) A \emph{road network} encoded in the Lanelet2~\cite{lanelets2} format that describes the ``drivable'' areas of the map. 
(ii) A set of \emph{obstacles} representing static and dynamic objects;
(iii) A set of \emph{planning problems} that encode driving tasks. Each planning problem includes the initial state of a vehicle (e.g., position, rotation) and conditions to be met (e.g., reaching a target position within a timeout) to complete the driving task successfully.
We use Kaufeld et al.'s platform to simulate CommonRoad scenarios and record the state of all the vehicles at each simulation step, which yields an execution trace.

\roadlogic assumes all the vehicles start from a standstill position and generates sequences of waypoints that must be reached within a specified time. We map these conditions to CommonRoad initial and goal states to obtain planning problems that FrenetiX solves during scenario execution by driving the simulated vehicles.
FrenetiX is an open-source motion planner for CommonRoad that implements a deterministic sampling-based planning algorithm. In a nutshell, FrenetiX computes an overall route from the initial state to the goal areas (i.e., the reference path) and cyclically (re)plans short-term trajectories that avoid collisions. FrenetiX identifies candidate short-term trajectories that are safe and physically feasible via rejection sampling, then selects the one that minimizes a user-defined cost function. Typical cost functions penalize trajectories that would result in unacceptable risk levels (safety), abrupt changes in direction and speed (comfort), or that are distant from the reference path.

\section{\roadlogic}
\label{sec:method}

In this section, we present \roadlogic, a framework for generating specification-compliant simulations from OpenSCENARIO DSL. 

\subsection{Overview}

\begin{figure*}[htb]
    \centering
    \includegraphics[width=\linewidth]{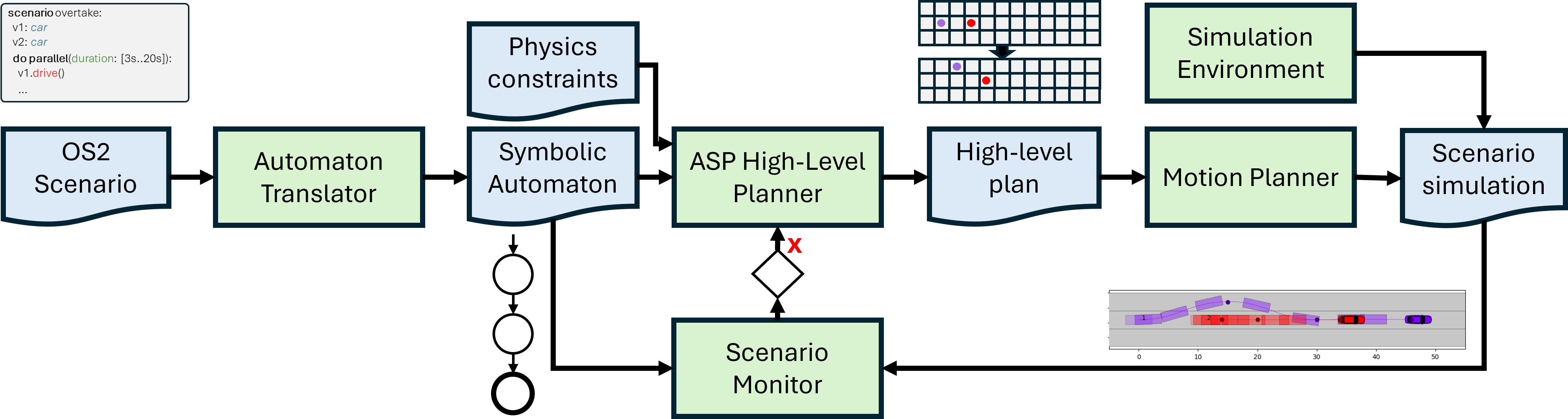}
    \caption{Overview of \roadlogic.}
    \Description{Overview diagram of the system.}
    \label{fig:pipeline}
\end{figure*}

Figure~\ref{fig:pipeline} shows an overview of \roadlogic. 
The OpenSCENARIO DSL specification is first translated into a symbolic automaton representation that encodes the sequential behavioral, relational, spatial and timing constraints of all actors.
The symbolic automaton representation of the abstract scenarios is encoded into a planning problem using the action language formalism. The encoding is extended with a simple set of general physics constraints to ensure the plausibility of the scenario instantiations.  
This high-level planning problem is solved using Answer Set Programming. A high-level plan is refined to a discrete sequence of waypoints for each actor.
The translation of the refined plan into a concrete simulation is performed using an advanced motion planner integrated into an AV simulation environment.
The concrete simulation is evaluated for compliance with respect to the original OpenSCENARIO DSL specification by monitoring the generated trace against the symbolic automaton representation of the abstract scenario.

\subsection{From OS2 to Symbolic Automata}
\label{sec:automaton}
The first step of \roadlogic consists in translating OS2 scenarios into an equivalent symbolic automata representation, in which each location represents a maneuver from the scenario and transitions are annotated with predicated encoding constraints (e.g. spatial relations between vehicles, their velocity) in the scenario. 

\begin{figure*}[t]
\centering
\begin{tikzpicture}[->, >=stealth, auto,
  nd/.style={draw, circle, minimum size=1.4em, inner sep=0pt, font=\scriptsize},
  acc/.style={draw, circle, minimum size=1.4em, inner sep=0pt, font=\scriptsize, double},
  wide/.style={draw, circle, minimum size=2.0em, inner sep=0pt, font=\tiny},
  wideacc/.style={draw, circle, minimum size=2.0em, inner sep=0pt, font=\tiny, double},
  sa/.style={draw, rectangle, rounded corners=1pt, font=\scriptsize,
             minimum width=0.8cm, minimum height=0.7cm},
  lbl/.style={draw=none, font=\tiny}]

\node[nd] (ai) at (0,0) {$i$};
\node[nd] (as) at (1.2,0) {$s$};
\node[acc] (af) at (2.4,0) {$f$};
\draw[->] (-0.6,0) -- (ai);
\path (ai) edge node[lbl,above] {$\gamma_1$} (as)
      (as) edge node[lbl,above] {$\gamma_2$} (af);
\node[draw=none] at (1.2,-1.0) {\scriptsize\textbf{(a)} Atomic};

\node[nd]  (bi)  at (4.0,0) {$i_1$};
\node[sa]  (bA1) at (5.2,0) {$\mathcal{A}_1$};
\node[sa]  (bA2) at (6.6,0) {$\mathcal{A}_2$};
\node[acc] (bf)  at (7.8,0) {$f_2$};
\draw[->] (3.4,0) -- (bi);
\draw[->] (bi)  -- (bA1);
\draw[->] (bA1) -- (bA2);
\draw[->] (bA2) -- (bf);
\node[draw=none] at (5.9,-1.0) {\scriptsize\textbf{(b)} Serial};

\node[wide]    (ci) at (9.4,0) {$(i_1,i_2)$};
\node[sa]      (cp) at (11.0,0) {$\mathcal{A}_1\!\times\!\mathcal{A}_2$};
\node[wideacc] (cf) at (12.7,0) {$(f_1,f_2)$};
\draw[->] (8.5,0) -- (ci);
\draw[->] (ci) -- (cp);
\draw[->] (cp) -- (cf);
\node[draw=none] at (11.0,-1.0) {\scriptsize\textbf{(c)} Parallel};

\node[nd]  (di)  at (14.4,0) {$i$};
\node[sa]  (dA1) at (15.7, 0.45) {$\mathcal{A}_1$};
\node[sa]  (dA2) at (15.7,-0.45) {$\mathcal{A}_2$};
\node[acc] (df)  at (17.0,0) {$f$};
\draw[->] (13.8,0) -- (di);
\draw[->] (di) -- (dA1);
\draw[->] (di) -- (dA2);
\draw[->] (dA1) -- (df);
\draw[->] (dA2) -- (df);
\node[draw=none] at (15.7,-1.0) {\scriptsize\textbf{(d)} One-of};

\end{tikzpicture}
\caption{Symbolic automaton translation rules for OS2 constructs.}
\label{fig:translation-rules}
\end{figure*}
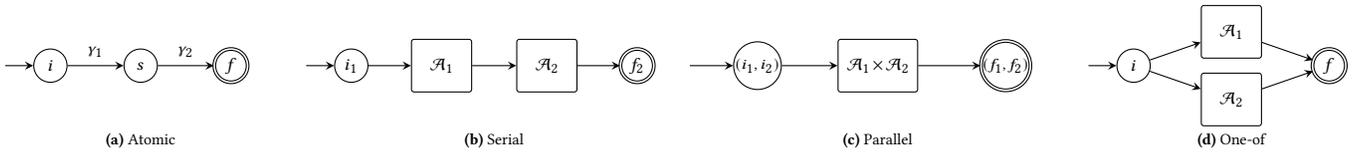

The translation $\AutT{\cdot}$ from an OS2 specification $\varphi$ into a symbolic automaton $\sa_{\varphi}$ is defined inductively below; Figure~\ref{fig:translation-rules} summarizes the corresponding constructions.

\subsubsection*{Atomic Maneuver:} The atomic behavior $\kwdriveshort{\gamma_1}{\gamma_2}$ is translated into a three-location automaton, with $i$ the initial, $f$ the final, and $s$ the intermediate location. The transition from $i$ to $s$ is labeled with the \emph{start} constraint $\gamma_1$, and the transition from $s$ to $f$ with the \emph{end} constraint $\gamma_2$: \\
$\AutT{\kwdriveshort{\gamma_1}{\gamma_2}} = (\{i,s, f\}, i, f, \{(i, \gamma_1,s),  (s, \gamma_2, f)\})$

For the composition rules, let 
$\sa_{1} = \AutT{\varphi_1} = (Q_1, i_1, f_1, \Delta_1)$ and
$\sa_{2} = \AutT{\varphi_2} = (Q_2, i_2, f_2, \Delta_2)$.

\subsubsection*{Serial Composition:} The serial composition $\varphi_{1} \cdot \varphi_2$ is translated by serially composing $\sa_1$ and $\sa_2$:
$\AutT{\varphi_1 \cdot \varphi_2} = (Q, i_1, f_2, \Delta)$, where:
    \begin{itemize}
        \item $Q = Q_1 \cup Q_2 \setminus \{f_1, i_2\}$
        \item $\Delta = \Delta_1 \cup \Delta_2' \cup \{(f_1, P, q) ~|~ (i_2, P, q) \in \Delta_2\}$
        \item $\Delta_2'$ excludes transitions involving $i_2$.
    \end{itemize}

\subsubsection*{Parallel Composition:} The parallel composition $\varphi_{1} \parallel \varphi_2$ is translated by taking the Cartesian product of $\sa_1$ and $\sa_2$: 
$\AutT{\varphi_1 \parallel \varphi_2} = (Q_1 \times Q_2, (i_1, i_2), (f_1, f_2), \Delta_{\parcomp})$, where $\Delta_{\parcomp}$ lifts transitions from $\Delta_1$ and $\Delta_2$ component-wise.

\subsubsection*{One-Of Composition:} The one-of composition $\varphi_1 \vee \varphi_2$ is translated by the union of $\sa_1$ and $\sa_2$: 
$\AutT{\varphi_1 \vee \varphi_2} = (Q, i, f, \Delta)$, where $Q = Q_1 \cup Q_2 \cup \{i, f\} \setminus \{i_1, f_1, i_2, f_2\}$ and
    \begin{align*}
        \Delta = {}&\{(i, P, q) \mid (i_1, P, q) \in \Delta_1\} \cup
                  \{(q, P, f) \mid (q, P, f_1) \in \Delta_1\} \\
              &\cup \{(i, P, q) \mid (i_2, P, q) \in \Delta_2\} \cup
                  \{(q, P, f) \mid (q, P, f_2) \in \Delta_2\} \\
              &\cup \Delta_1' \cup \Delta_2',
    \end{align*}
    with $\Delta_1'$ and $\Delta_2'$ excluding transitions involving $i_1, f_1, i_2, f_2$.

\begin{example}[Symbolic Automaton of the Overtake Scenario] Figure~\ref{fig:translation-graph} depicts the symbolic automaton generated from the OS2 overtake scenario from Listing~\ref{lst:os2}. 

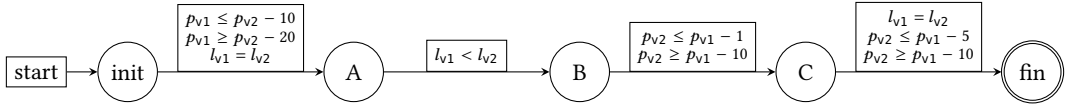
\begin{figure*}[ht]
    \centering
    \begin{tikzpicture}[->, >=stealth, auto, every node/.style={draw, minimum size=1.2em}, baseline=(init.base), node distance=3.0cm and 2.2cm]
        \node[state, initial] (init) {init};
        \node[state] (A) [right=of init] {A};
        \node[state] (B) [right=of A] {B};
        \node[state] (C) [right=of B] {C};
        \node[state, accepting] (fin) [right=of C] {fin};
    
        \path
            (init) edge node[above, font=\scriptsize, align=center] {$p_{\ego} \leq p_{\car} - 10$\\$p_{\ego} \geq p_{\car} - 20$\\$l_{\ego} = l_{\car}$} (A)
            (A) edge node[above, font=\scriptsize] {$l_{\ego} < l_{\car}$} (B)
            (B) edge node[above, font=\scriptsize, align=center] {$p_{\car} \leq p_{\ego} - 1$\\$p_{\car} \geq p_{\ego} - 10$} (C)
            (C) edge node[above, font=\scriptsize, align=center] {$l_{\ego} = l_{\car}$\\$p_{\car} \leq p_{\ego} - 5$\\$p_{\car} \geq p_{\ego} - 10$} (fin);
    \end{tikzpicture}
    \caption{Symbolic automaton derived from the overtake scenario}
    \Description{Symbolic automaton of the running example.}
    \label{fig:translation-graph}
\end{figure*}
       
\end{example}

The use of the symbolic automaton representation of the OS2 scenario is twofold: to facilitate the ASP planning problem definition and to monitor whether the concrete simulation is compliant with the original scenario.

\subsection{ASP Encoding of Scenario Constraints}
\label{sec:asp-encoding}

We encode the instantiation of OS2 scenarios as a discrete-time planning problem in ASP.
In the planning view, the objective is to find a sequence of actions that leads from an initial state to a goal state while satisfying all scenario and domain constraints.
Following the conventions in ASP planning~\cite{actionsurvey_2023}, a planning problem is represented by: 
(i) a set of \emph{fluents} describing the possible states of the world,
(ii) a set of \emph{actions} with their preconditions and effects, and
(iii) a set of \emph{transition rules} that relate successive time steps.
Solving the ASP program yields one or more \emph{answer sets}, each corresponding to an abstract plan.

\subsubsection*{Incremental solving.} 
To efficiently compute plans in evolving domains, we 
employ 
\emph{multi-shot solving}~\cite{DBLP:journals/tplp/GebserKKS19}.
Instead of repeatedly re-grounding and re-solving the entire program for increasing horizons, multi-shot solving allows adding or retracting rules and incrementally extending the temporal scope of the problem.
We structure our ASP program into three subprograms:
\texttt{\#program base}, which contains static domain facts and constants;
\texttt{\#program step(t)}, which defines the dynamic transition rules per time step; and
\texttt{\#program check(t)}, which expresses goal satisfaction.
At each iteration, the solver incrementally advances the time step~$t$, adding the corresponding rules and checking whether all goal constraints are met.
This approach combines the declarativity of ASP with the iterative refinement commonly found in planning frameworks.

\subsubsection*{Discretization of time and space.}
We abstract the continuous driving environment by discretizing both space and time into a grid-based representation.
Time progresses in integer steps~$t = 0, 1, 2, \ldots$, each corresponding to a pre-defined unit of real time (e.g. one second).
The road network is modeled as a two-dimensional grid: the longitudinal $x$-axis corresponds to distance in meters (\emph{segments}), and the lateral $y$-axis corresponds to lane identifier integers.
Thus, a cell $(S,L)$ represents a rectangle of one meter length and one lane width.
The ego vehicle starts at position $(0,0)$ at time~0.
Vehicle actors move along the grid according to longitudinal and lateral velocities, expressed as integers in segments per step and lanes per step, respectively.
We restrict lateral motion to at most one lane per time step (\texttt{velocity\_lat~$\in\{-1, 0, 1\}$}), while the longitudinal velocity is non-negative.

\subsubsection*{State representation.}
We represent the discrete state of each actor (vehicle) with the fluent \texttt{state(V, loc(S,L), dyn(Vel\_lon, Vel\_lat))}, indicating that vehicle~$V$ is located at segment~$S$ and lane~$L$ with longitudinal and lateral velocities \texttt{Vel\_lon} and \texttt{Vel\_lat}.
In planning notation, we express that a fluent holds at time~$T$ as:
\begin{verbatim}
holds(state(V, loc(S,L), dyn(Vel_lon, Vel_lat)), T).
\end{verbatim}
This representation allows compact reasoning about the evolution of all vehicles across time and space.

\subsubsection*{Action representation.}
Changes in vehicle dynamics are modeled as actions of the form:
\begin{verbatim}
    change_dynamics(V, Delta_Vel_lon, Vel_lat).
\end{verbatim}
Each action updates the longitudinal velocity by \texttt{Delta\_Vel\_lon} (representing instantaneous acceleration) and sets a new lateral velocity~\texttt{Vel\_lat}.
Exactly one such action occurs for each vehicle at every time step:
\begin{verbatim}
{occurs(change_dynamics(V, Delta_Vel_lon, Vel_lat), t): 
acceleration_lon(Delta_Vel_lon), velocity_lat(Vel_lat)} 
= 1 :- vehicle(V).
\end{verbatim}
This compact choice rule~\cite{DBLP:journals/cacm/BrewkaET11, DBLP:series/synthesis/2012Gebser} encodes the non-deterministic selection of a single action from the available options.

\subsubsection*{State transitions.}
The new state at time~$t$ is derived deterministically from the previous state and the applied actions as follows:
\begin{verbatim}
holds(state(V, loc(S,L), dyn(Vel_lon,Vel_lat)), t) :-
    holds(state(V, loc(S_prev,L_prev), 
          dyn(Vel_lon_prev,Vel_lat_prev)), t-1),
    occurs(change_dynamics(V, Delta_lon, Vel_lat), t),
    L = L_prev + Vel_lat_prev,
    S = S_prev + Vel_lon_prev,
    Vel_lon = Vel_lon_prev + Delta_Vel_lon,
    velocity_lon(Vel_lon).
\end{verbatim}
This transition rule propagates the vehicle position according to its prior velocity, updated by the acceleration at step~$t$.

\subsubsection*{Symbolic automaton constraints.}
The symbolic automaton derived from the OS2 specification (Section~\ref{sec:automaton}) defines the logical structure of the scenario through its nodes~$s_i$ and edges~$e_i = (s_{i-1}, \gamma_i, s_i)$, where each label~$\gamma_i$ represents a conjunction of symbolic predicates that must hold to enable the transition.
For each $\gamma_i$, we generate a goal rule that enforces its satisfaction at some time~$T_i \le t$:
\begin{verbatim}
    goal_i(t) :- query(t), T <= t, gamma_i(T).
\end{verbatim}
The sequential satisfaction of all such subgoals forms the overall scenario goal:
\begin{verbatim}
goal(T1, ..., Tn, t) :- goal_1(T1), ..., goal_n(Tn), 
                        0 <= T1 <= ... <= Tn <= t.
\end{verbatim}
These goal rules are placed in the \texttt{check(t)} subprogram, ensuring that the multi-shot solver only terminates once all required conditions along the automaton path are satisfied.

\subsubsection*{Domain constraints.} We combine the constraints explicitly defined in the OS2 scenario with additional domain constraints to ensure physical and behavioral realism:
(1) vehicles must not stop completely after initialization,
(2) longitudinal and lateral velocities remain within specified bounds,
(3) vehicles stay within road limits,
(4) lane changes cannot occur in consecutive steps, and
(5) collisions are forbidden.
For instance, to prevent immediate successive lane changes, we define a blocking duration~\texttt{change\_dur} after a lane change and enforce that vehicles remain on the new lane during this interval.
Similarly, safety distances are maintained by prohibiting configurations in which two vehicles occupy the same or adjacent segments within a minimum longitudinal gap.

\begin{figure}[h]
    \centering
    \includegraphics[width=\columnwidth]{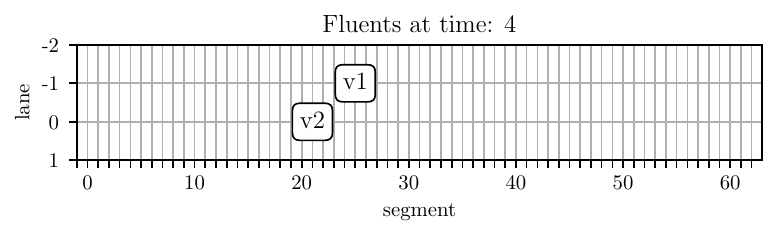}\\
    \includegraphics[width=\columnwidth]{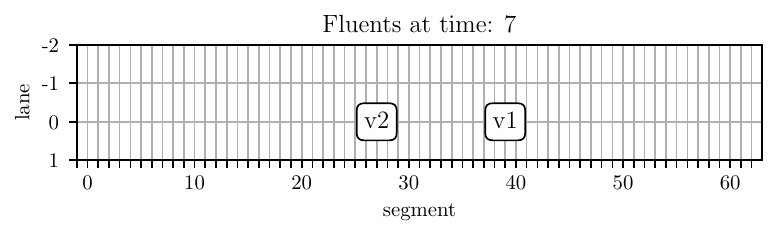}\\
    \includegraphics[width=\columnwidth]{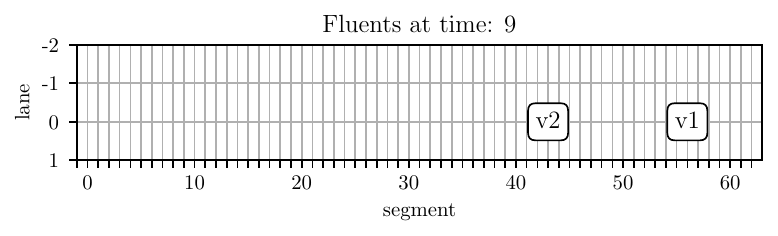}
    \caption{Selected frames from the ASP model output of the running example.}
    \Description{The points in time when the three goals are fulfilled.}
    \label{fig:asp-model-frames}
\end{figure}

\begin{example}[ASP Encoding of the Overtake Scenario]
Listing~\ref{lst:overtake-instance} shows an excerpt of the ASP encoding for the overtake scenario used in our running example.
The \texttt{init} rules encode the guard on edge (\texttt{init}, A) in Figure~\ref{fig:translation-graph}, placing \ego behind \car on the same lane.
The goal rules describe three sequential goals, corresponding to edges (A,B), (B,C), and (C,\texttt{fin}) in Figure~\ref{fig:translation-graph}:
(i) the vehicle \ego is left of the other car \car,
(ii) it reaches a longitudinal offset ahead of \car, and
(iii) it returns to the same lane in front of it.
The solver incrementally extends the horizon until all goals are achieved.
Figure~\ref{fig:asp-model-frames} visualizes the corresponding time steps in which these conditions hold.
\end{example}

\begin{aspbox}[title={Listing~\ref{sec:asp-encoding}: ASP encoding of a goal sequence}, label={lst:overtake-instance}]
% INITIAL SETUP FACTS:

#program base.
init(state(@\aspego@,loc(0,0),dyn(1,0))).
1 { init(state(@\aspcar@, loc(S1, L1), dyn(1,0))) : segment(S1), lane(L1), -20 <= (S2 - S1) <= -10, L2 = L1 } 1 
	:- init(state(@\aspego@, loc(S2, L2), dyn(1,0))).
vehicle(@\aspego@).
vehicle(@\aspcar@).

#program check(t).
% GOAL RULES:
goal1(t) :- holds(state(@\aspego@,loc(S1,L1),dyn(_,_)), t), holds(state(@\aspcar@,loc(S2,L2),dyn(_,_)), t), L1 < L2.
goal2(t) :- holds(state(@\aspego@,loc(S1,L1), dyn(_,_)), t), holds(state(@\aspcar@,loc(S2,L2),dyn(_,_)), t), 1 <= (S1 - S2) <= 10.
goal3(t) :- holds(state(@\aspego@,loc(S1,L1),dyn(_,_)), t), holds(state(@\aspcar@,loc(S2,L2),dyn(_,_)),t), 5<=(S1 - S2)<=10, L1=L2.
% GOAL ASSERTION:
goal(T1,T2,T3,t) :-  goal1(T1), goal2(T2), goal3(T3), 0 <= T1<=T2<=T3 <= t, t > T3 + 1.
:- query(t), not goal(_,_,_,t).
\end{aspbox}

\subsubsection*{From answer sets to concrete scenarios.}
Each computed answer set represents a high-level discrete plan, containing facts of the form:
\begin{verbatim}
holds(state(V, loc(S,L), dyn(Vel_lon, Vel_lat)), T).
\end{verbatim}
We extract these fluents to define the initial state and a sequence of waypoints which serve as the basis for generating concrete scenarios for each vehicle.
In this paper, we illustrate the application of \roadlogic by generating concrete scenarios in CommonRoad; however, the approach is generic and can be adapted to generate scenarios in other simulators.
Therefore, we encode information about initial, intermediate, and final expected states into CommonRoad planning problems (see~Section~\ref{sec:commonroad}) but force all vehicles to start from a standstill position 
in compliance with state-of-the-art driving simulators, such as CARLA~\cite{Dosovitskiy17} and BeamNG.tech~\cite{beamng_tech}. 

As described in the next section, we simulate the concrete CommonRoad scenarios using Kaufeld et al.'s platform and the FrenetiX motion planner.

\subsection{Concrete Scenario Execution and Monitoring}

The ASP solver determines \emph{what} must happen at the symbolic level, i.e., the ordering of maneuvers and relational constraints between actors, while Frenetix determines \emph{how} these constraints are realized in continuous space under vehicle dynamics and road geometry. Multiple trajectories may satisfy the same high-level plan; \roadlogic controls concretization by enforcing ASP-derived intermediate goals in sequence and by rejecting executions that violate the symbolic automaton during monitoring.

\roadlogic generates scenarios containing multiple intermediate goals. However, neither Kaufeld et al.'s simulation platform
nor the FrenetiX motion planner support this setup.

Specifically, Kaufeld et al.'s platform assumes that driving agents must achieve a single goal; therefore, 
it removes them from the simulation as soon as they reach the target position.
Likewise, the FrenetiX motion planner, which also assumes there is a single target goal, 
cannot plan a reference path that traverses all the intermediate locations before reaching the final one.

To solve these limitations, we extended the original implementation of Kaufeld et al.'s platform to keep track of intermediate goals and remove the agents only when they reach their final goal. We extended FrenetiX by allowing the driving agents to dynamically change their (current) goals, such that agents will recompute the reference path after reaching an intermediate goal. 
We used the updated platform and motion planner to solve the (multi-goal) planning problems derived from the ASP high-level plans, effectively using the simulations to refine the high-level plans into physically feasible trajectories that respect vehicle dynamics, road geometry, and scenario specifications. 

After executing a scenario, \roadlogic collects the execution traces and monitors their compliance with the symbolic automaton specifications, using the algebraic runtime verification~\cite{DBLP:journals/tcad/JaksicBGN18} framework. Additionally, \roadlogic checks that the vehicles reached the goals within the allotted time and that all the vehicles completed the whole simulation, ensuring that also the global invariants of the execution hold. If the generated execution trace does not comply with the symbolic automaton specification, and hence to the intended OS2 scenario, the concrete instantiation is rejected and a new plan is generated.

\subsection{Implementation}

\roadlogic\ is implemented in Python~3.10 and released under the permissive BSD~3-Clause license. 
It includes an ANTLR4-based parser for \osTwo\ scenarios, using the official grammar from the community project \texttt{py-osc2}~\cite{pyosc2}. 
The high-level planning engine relies on the Potassco \clingo\ ASP solver~\cite{DBLP:journals/corr/GebserKKS14}, 
while the low-level motion planning is performed by the Frenetix planner~\cite{Raineretal2024Frenetix} 
within the CommonRoad simulation framework~\cite{commonroad2017}.
The complete source code and experiment scripts are publicly available at 
\url{https://github.com/figlerg/roadlogic/tree/artifact-hscc26}. 
Animated visualizations of the executions shown in Figures~\ref{fig:working-example}, \ref{fig:asp-model-frames}, \ref{fig:var}, and~\ref{fig:unknown} are available in the replication package under \texttt{paper/animations}.

\section{Evaluation}
\label{sec:evaluation}

In this section we present the evaluation results. We first formulate our research questions, describe the experimental setup and then perform the evaluation that allows us to respond to the questions.

\subsection{Research Questions}

We seek to answer the following research questions:

\begin{researchquestion}{RQ1}{Effective Scenario Generation}
\emph{Can \roadlogic generate scenarios that result in realistic executions and satisfy the declarative \osTwo specifications in an acceptable time?}

\textbf{Rationale}:
Addressing this research question determines whether \roadlogic reliably and efficiently bridges the gap between declarative scenario specifications and executable simulations.

\textbf{Metrics}:
To demonstrate that \roadlogic generates scenarios that satisfy the specification, we verify that the monitors return a positive verdict after checking each scenario execution trace. We also verify this result using visual inspection.
We log the time to complete each step of \roadlogic pipeline (see Section~\ref{sec:method}) to assess overall performance and the impact of each activity. 
\end{researchquestion}

\begin{researchquestion}{RQ2}{Generated Scenario Diversity}
\emph{Can \roadlogic generate quantifiably different scenarios
from the same \osTwo specifications?}

\textbf{Rationale}: 
Systematically generating diverse scenarios might stress different behaviors and possibly expose more bugs, thus improving testing effectiveness.
Doing so while ensuring that all scenarios meet developers' expectations ensures they fulfill the intended purpose, thereby achieving high test quality.

\textbf{Metrics}:
To assess how different the generated scenarios are, we implemented a scenario similarity metric based on occupancy grid~\cite{DBLP:conf/itsc/BraunFRRPSS23}.
In a nutshell, we measure the similarity between scenarios as the Jaccard Index of their observed occupancy of the grid:
\begin{equation}
\text{similarity}(S_1,S_2) = \frac{|OG_1 \cap OG_2|}{|OG_1 \cup OG_2|}
\end{equation}
where OG$_i$ is the set of cells in the grid representation (see~Section~\ref{sec:asp-encoding}) occupied by any vehicle in scenario S$_i$ at any time.
Similarity values close to $1$ indicate vehicles covered similar ground; conversely, values close to $0$ suggest that the vehicles followed different trajectories.
\end{researchquestion}

\subsection{Experimental Setup}
We evaluate \roadlogic on a suite of \numscenarios \osTwo specifications covering various scenarios and variations:
\begin{enumerate*}
\item \emph{Follow} is a car-following scenario that involves two vehicles, lead and tail, driving in the same direction; the tail must follow the lead's trajectory.
\item \emph{Overtake} is an overtaking scenario that involves two vehicles, lead and tail, driving in the same direction in the same lane; the tail drives faster than the lead, thus it must switch lane and pass it to avoid colliding.
\item \emph{Overtake~+~3rd~actor}, \item \emph{Overtake~+~fixed~lane}, and 
\item \emph{Overtake~+~obstacle} are variations of the basic overtaking scenario and involve a third vehicle (moving and standing) or force the lead to keep its lane. 
\item \emph{Change~lane} is a scenario where one vehicle changes lane while the other drives freely.
\item \emph{Dodge~obstacle} involves a vehicle traveling in a lane that must avoid a second vehicle parked on the same lane.
\end{enumerate*}

We evaluate two different strategies of using \roadlogic: (1) \emph{base} strategy in which all responsibility for generating different concrete scenarios is given to the ASP planner, and (2) \emph{refined} strategy in which the input 
parameters (e.g., initial actor distances and dynamic constraints) are sampled from given ranges and the ASP planner generates plans with the sampled parameters instead.

To address RQ1, we use \roadlogic with the base strategy. In this \emph{base} experiment, we generate $10$ concrete scenarios for each of the \numscenarios \osTwo specifications. 

To address RQ2, we use \roadlogic with the refined strategy and randomly sample the scenario input parameters, generating different ASP planning problems. 
In this 
experiment, we generate $50$ concrete scenarios for each of the \numscenarios \osTwo specifications.

\subsection{RQ1. Effective Scenario Generation.}
We evaluate \roadlogic using the base strategy. For each scenario, we generate 10 concrete simulations. We use the scenario monitors to check the compliance of the simulations to the abstract scenarios. The evaluation shows that all the instantiated simulations from all the scenarios conform to their respective scenario specification. 

\begin{figure}
    \centering
    \includegraphics[width=\linewidth]{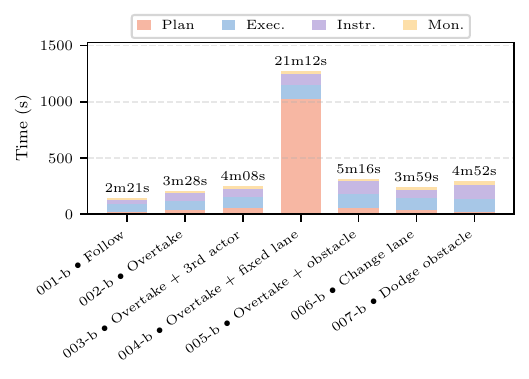}
    \caption{Computation time of base pipeline on the scenario benchmark.}
    \Description{}
    \label{fig:vanillacompute}
\end{figure}

Figure~\ref{fig:vanillacompute} shows the computation times for generating scenarios using the base strategy. For each scenario, we split the total computation time across the main (planning, execution, instrumentation, monitoring) tasks.
We see that most benchmark specifications complete the full scenario generation pipeline within minutes. We also observe that the computation time is sensitive to the minimum time horizon needed for a satisfying plan. In all scenarios except one, the execution (i.e. simulation in the CommonRoad framework) is the activity that dominates the overall scenario generation time.  One major exception is Experiment 4, in which the high-level ASP planning presents a clear computational bottleneck. This scenario requires a sequence of complex maneuvers (including two consecutive changes of the lane) and requires a longer time horizon for completion, thus increasing the complexity of the ASP planning problem. 

\begin{graybox}
\textbf{RQ1:} \roadlogic is effective in generating concrete executions from abstract OS2 scenarios with reasonable cost in most of the cases.
\end{graybox}

\subsection{RQ2. Scenario Diversity}

We evaluate the scenario diversity generated by \roadlogic with both the base and the refined strategy. 

\begin{figure}[t]
    \centering
    \begin{minipage}{\columnwidth}
        \centering

        \begin{subfigure}[t]{0.24\linewidth}
            \centering
            \includegraphics[width=\linewidth]{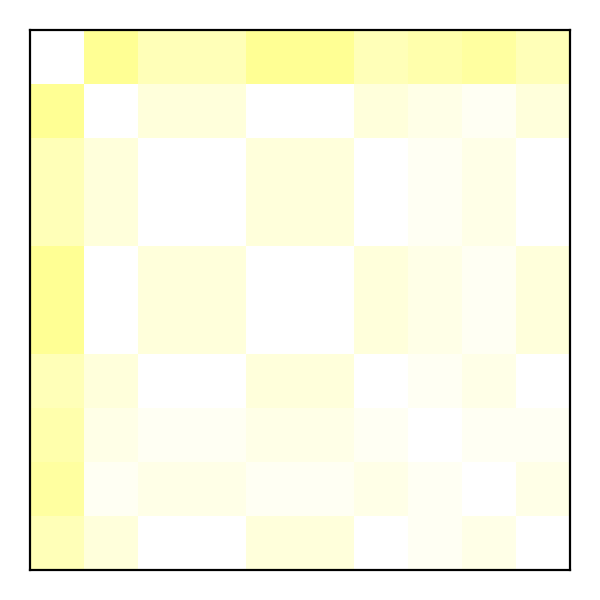}
            \caption*{001-b}
        \end{subfigure}\hfill
        \begin{subfigure}[t]{0.24\linewidth}
            \centering
            \includegraphics[width=\linewidth]{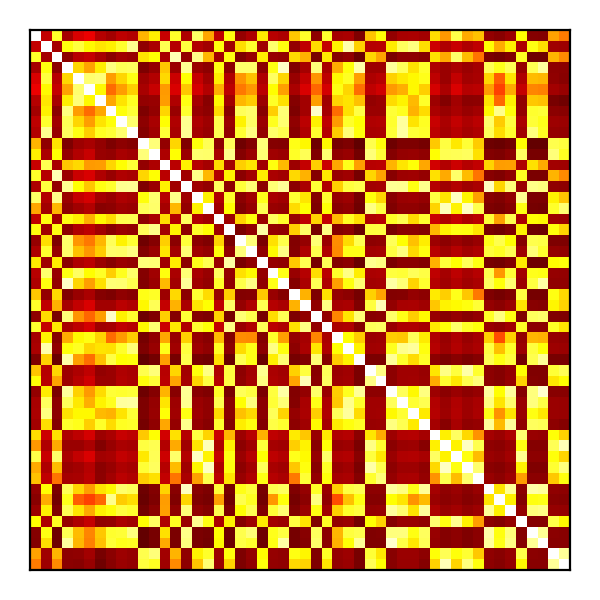}
            \caption*{001-r}
        \end{subfigure}\hfill
        \begin{subfigure}[t]{0.24\linewidth}
            \centering
            \includegraphics[width=\linewidth]{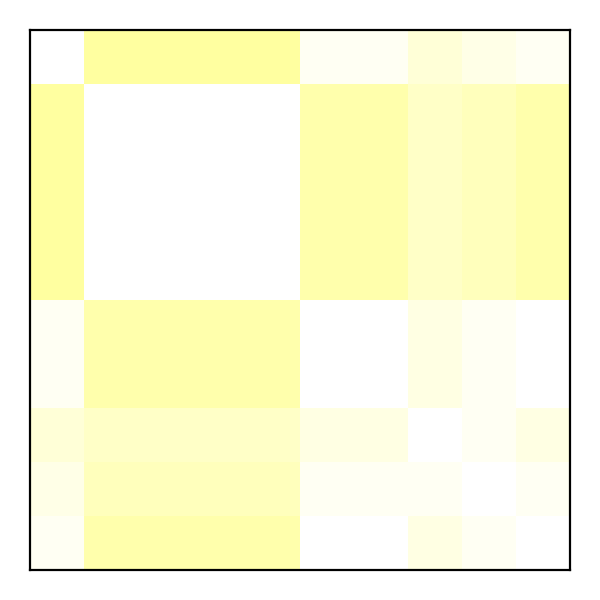}
            \caption*{002-b}
        \end{subfigure}\hfill
        \begin{subfigure}[t]{0.24\linewidth}
            \centering
            \includegraphics[width=\linewidth]{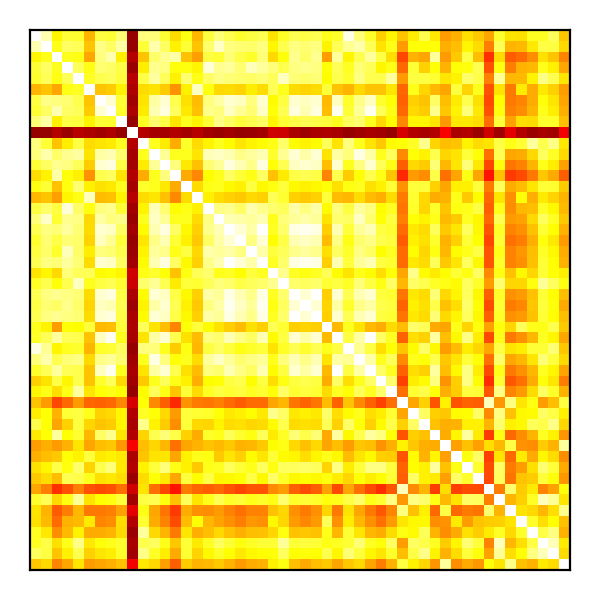}
            \caption*{002-r}
        \end{subfigure}

        \vspace{0.3em}

        \begin{subfigure}[t]{0.24\linewidth}
            \centering
            \includegraphics[width=\linewidth]{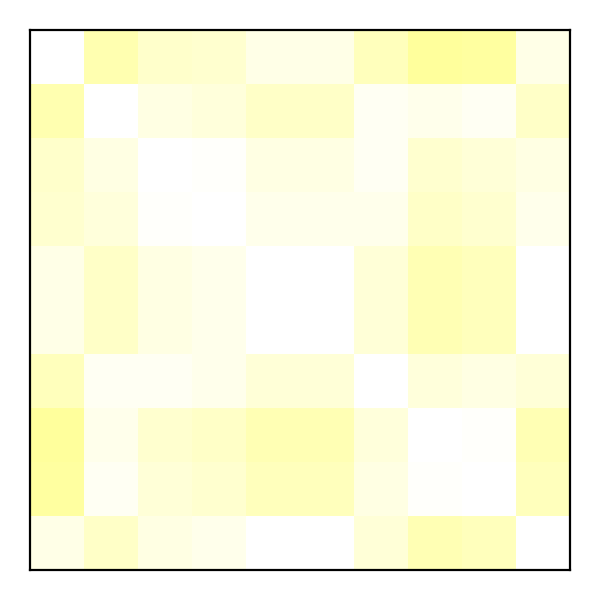}
            \caption*{003-b}
        \end{subfigure}\hfill
        \begin{subfigure}[t]{0.24\linewidth}
            \centering
            \includegraphics[width=\linewidth]{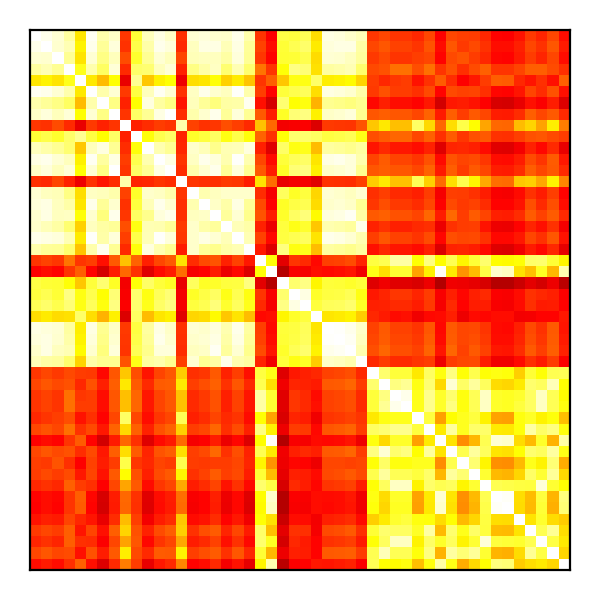}
            \caption*{003-r}
        \end{subfigure}\hfill
        \begin{subfigure}[t]{0.24\linewidth}
            \centering
            \includegraphics[width=\linewidth]{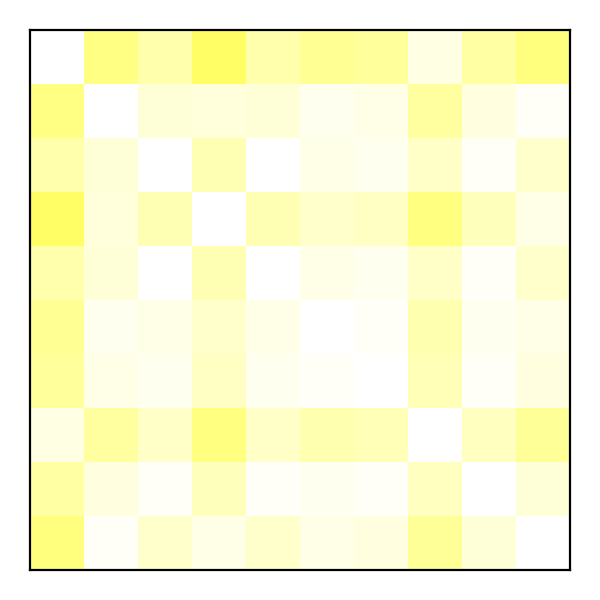}
            \caption*{004-b}
        \end{subfigure}\hfill
        \begin{subfigure}[t]{0.24\linewidth}
            \centering
            \includegraphics[width=\linewidth]{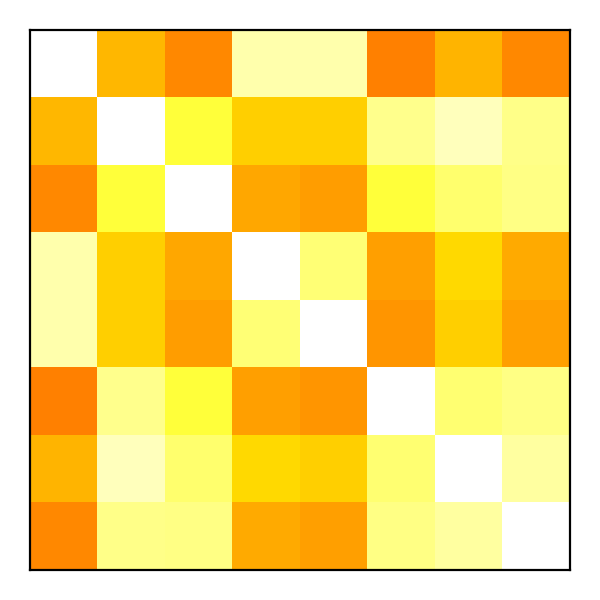}
            \caption*{004-r}
        \end{subfigure}

        \vspace{0.3em}

        \begin{subfigure}[t]{0.24\linewidth}
            \centering
            \includegraphics[width=\linewidth]{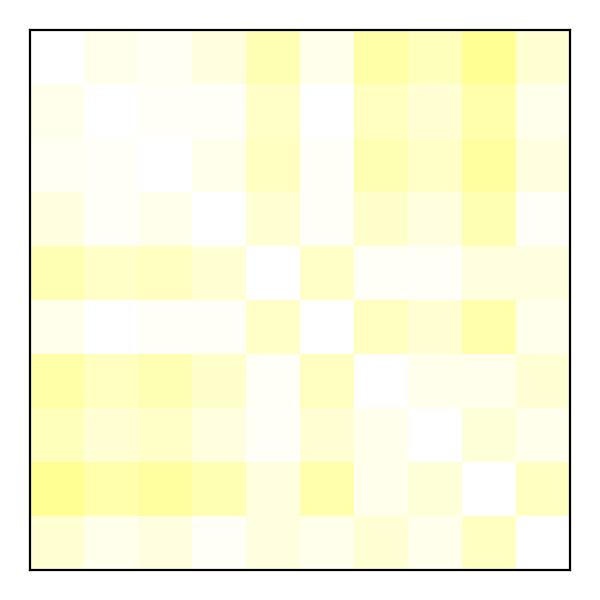}
            \caption*{005-b}
        \end{subfigure}\hfill
        \begin{subfigure}[t]{0.24\linewidth}
            \centering
            \includegraphics[width=\linewidth]{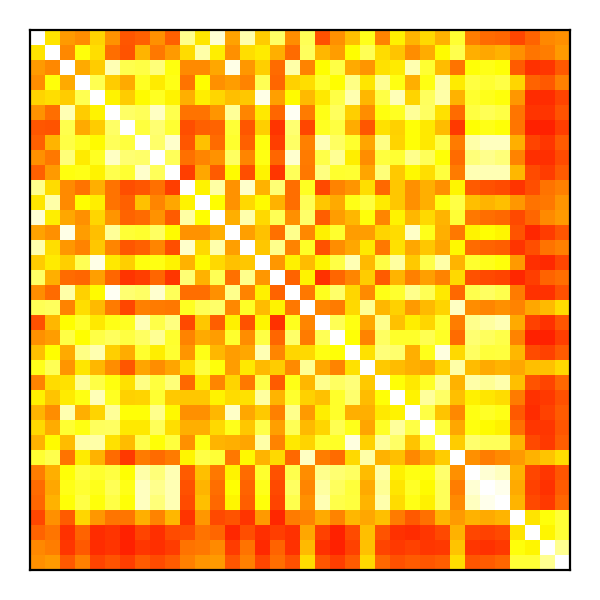}
            \caption*{005-r}
        \end{subfigure}\hfill
        \begin{subfigure}[t]{0.24\linewidth}
            \centering
            \includegraphics[width=\linewidth]{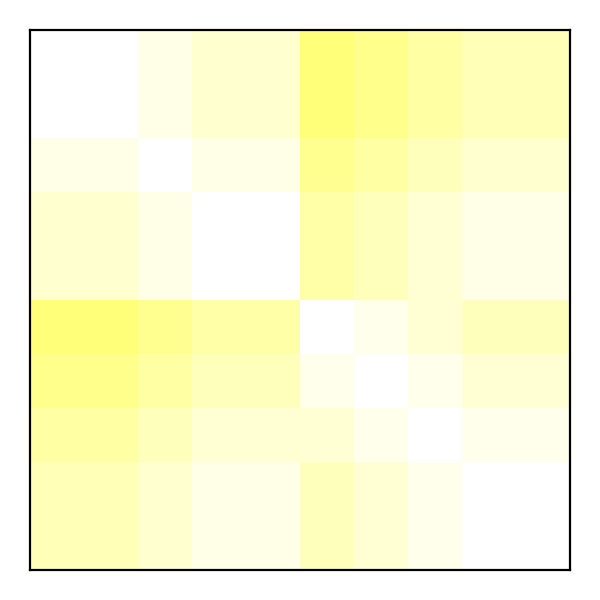}
            \caption*{006-b}
        \end{subfigure}\hfill
        \begin{subfigure}[t]{0.24\linewidth}
            \centering
            \includegraphics[width=\linewidth]{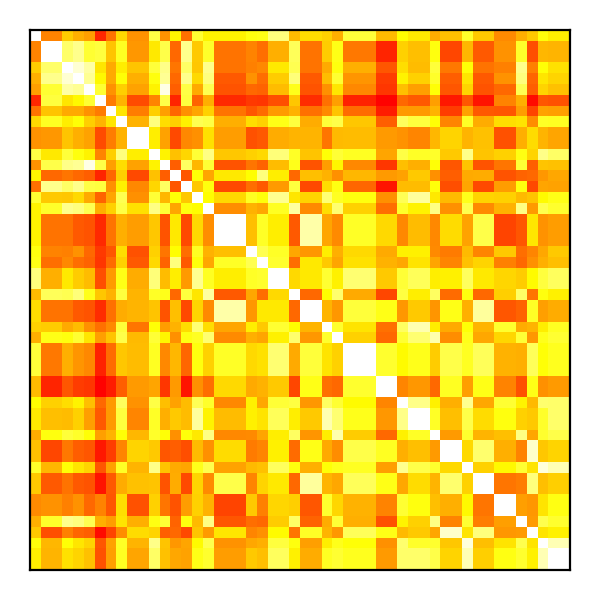}
            \caption*{006-r}
        \end{subfigure}

        \vspace{0.3em}

        \begin{subfigure}[t]{0.5\linewidth}
        \vspace{-1.2cm}\includegraphics[width=\linewidth]{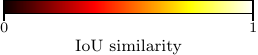}\\[0.4em]
        \end{subfigure}
        \begin{subfigure}[t]{0.24\linewidth}
            \centering
            \includegraphics[width=\linewidth]{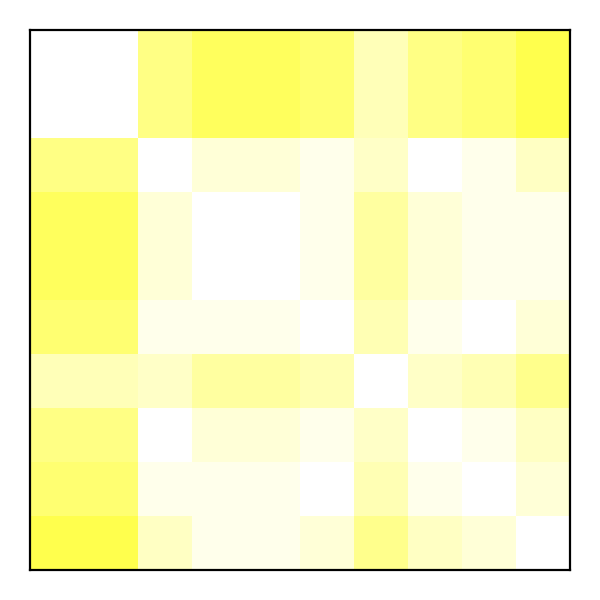}
            \caption*{007-b}
        \end{subfigure}
        \begin{subfigure}[t]{0.24\linewidth}
            \centering
            \includegraphics[width=\linewidth]{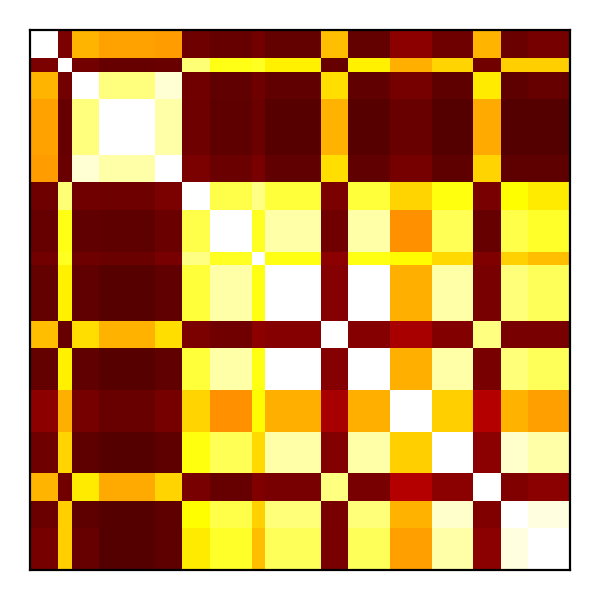}
            \caption*{007-r}
        \end{subfigure}

    \end{minipage}

    \caption{Similarity heatmaps for all experiments. Each pair (x-b, x-r) corresponds to one scenario under the base strategy and one under refined settings. All heatmaps share the color scale shown in the bottom left.}
    \Description{}
    \label{fig:similarity-all}
\end{figure}

Figure~\ref{fig:similarity-all} shows the variability results for both strategies, with each plot having the label X-Y, where X denotes the scenario number and Y whether the experiment was done using the base or the refined strategy. We first observe that the base strategy results in scenario instantiations with little diversity.  
The reason for this lack of diversity is that we use the incremental mode of the \clingo solver. In the incremental mode, the plan generation is done doing the depth-first search. In essence, a new plan is typically created by varying the last step of the old plan, thus considerably limiting the variation in the prefix of the generated plans.

\begin{figure}[ht]
\centering
\includegraphics[width=\linewidth]{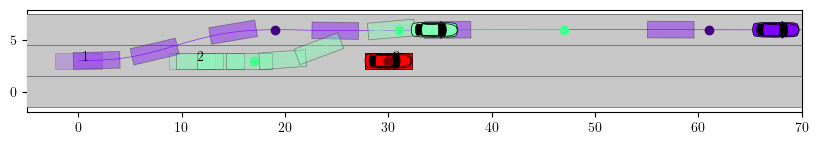}\\[1ex]
\includegraphics[width=\linewidth]{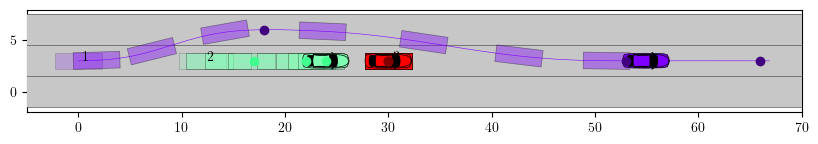}\\[1ex]
\includegraphics[width=\linewidth]{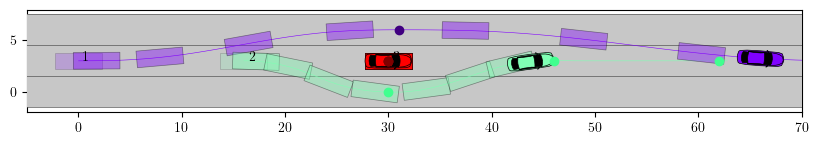}
\caption{Three concrete simulations generated from Scenario 5 using the refined strategy.}
\Description{}
\label{fig:var}
\end{figure}

In contrast, the refined strategy results in much higher variation of the generated scenario instances, as shown in Figure~\ref{fig:similarity-all}. We illustrate this variability by depicting three different instantiations of the Scenario 5 using the refined strategy in Figure~\ref{fig:var}. 
To enforce input parameter variation, every sample is encoded as a different ASP planning problem. 
We observe clustering in some scenarios (especially scenarios 3 and 7). This can be explained by an implementation detail of the parameter variation -- the parameter variation proceeds in alphanumerical order of the sampled parameters, meaning that these clusters likely are setups where the parameters flip between two sets of qualitatively different plans.
We also observe that the parameter variation impacts the diversity of under-constrained scenarios (e.g. scenarios 1 or 7) more than those that have tighter qualitative constraints (e.g. scenario 2).  This is not surprising since over-constrained scenarios leave less freedom for diversity.

\begin{figure}
    \centering
    \includegraphics[width=\linewidth]{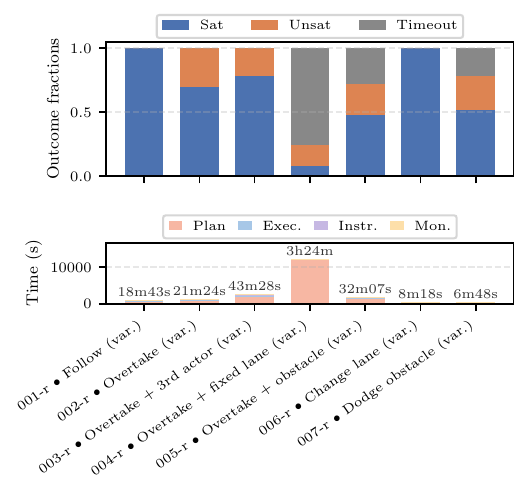}
    \caption{Success rate and computation time of pipeline with sampled parameters on the scenario benchmark.}
    \Description{}
    \label{fig:diversitysuccess}
\end{figure}

The enforcement of the parameter variations, resulting in the high variability, also comes at a cost. 
First, the scenario instantiation success rate is not anymore 100\% as for the base strategy. This is reflected in Figure~\ref{fig:diversitysuccess} (top). It shows that with the refined strategy, \roadlogic still generates many successful scenario instantiations (blue bar), but also sometimes simulations that do not conform to the scenario specification (orange bar) or times out without generating a simulation (grey bar). Scenario 4 seems especially hard to resolve using the refined strategy, where we observe many timeouts. 
We also see in Figure~\ref{fig:diversitysuccess} (bottom) that the refined strategy increases computation times compared to the base strategy (Figure~\ref{fig:vanillacompute}).

\begin{graybox}
\textbf{RQ2:} \roadlogic can instantiate abstract OS2 scenarios with high-variability, especially when input parameters are sampled from their applicable ranges.
\end{graybox}

\subsection{Threats to Validity}

\subsubsection*{Internal validity:} The restriction in this paper to a fragment of OS2 can represent a threat to validity. To mitigate this threat, we selected the representative subset of OS2 that is sufficient to demonstrate the feasibility and the effectiveness of the approach.

The depth-first search of the ASP solver tends to favor structurally similar solutions and hence can introduce the search bias. This bias can artificially limit scenario diversity and prevent exploring hard-to-reach parts of the search space. We address this threat by defining an alternative refined strategy that enforces more random input parameter choices.

\subsubsection*{External Validity:} Like many other evaluations, our study faces an external validity concern regarding the generalizability of the results. To address this, we experimented with seven different scenarios and two simulation environments with different physics properties and levels of accuracy. The collected evidence is consistent across scenarios. 

The evaluation focuses on scenarios with a small number of actors to keep the symbolic planning and monitoring behavior interpretable in this initial study. This is not a conceptual limitation of \roadlogic; larger multi-agent scenarios are primarily limited by ASP planning complexity and simulation cost, and constitute an important direction for future work.

\subsubsection*{Construct Validity:} The diversity evaluation relies on an occupancy-grid similarity metric. This metric may overemphasize certain geometric differences while under-representing distinct behavioral phenomena (e.g., different speed profiles or interactions). Thus, conclusions about diversity may not be fully complete and may depend on the chosen metric.

\section{Related Work}
\label{sec:related}

OpenSCENARIO\footnote{\url{https://www.asam.net/standards/detail/openscenario-xml/}} is a family of standard description languages for automated driving systems (ADS) scenarios.  Version XML (also known as 1.x)~\cite{asam_openscenario_userguide_v1_0, asam_openscenario_model_doc_v1_0} consists of an XML-based document that specifies step-by-step, using a procedural and imperative language, the behaviour, triggers, and manoeuvres of ADS.  Although this version is suitable for well-defined, concrete scenarios that a simulation engine can execute directly, this standard lacks the flexibility to specify abstract scenarios that describe the intent or what the scenario should look like, leaving simulation engines to determine how to execute them.  In contrast,  OpenSCENARIO DSL (also known as 2.0) (OS2)\footnote{\url{https://www.asam.net/standards/detail/openscenario/v200/}}  provides an object-oriented declarative domain specification language (DSL) that can describe more complex dynamic scenarios more abstractly by specifying the goal or intent of the scenarios rather than a fixed and concrete sequence of events, allowing also parametric and probabilistic variations. It is important to note that apart from the common OpenSCENARIO name, the XML and DSL variants have little in common, the former covering the specification of logical and the latter the abstract scenarios.  \textbf{\roadlogic is built on top of OS2}, enabling the possibility to concretize abstract OS2 scenarios into executable and monitorable ones through planning using ASP and compliant-checking using symbolic automata. 

Several survey papers provide useful taxonomies~\cite{zhong2021survey,Tang2023} about \emph{Scenario-based testing} or focus on automating the generation process~\cite{gao2025,Zhang2021} and scenario analysis.  In particular, Zhang et al.\cite{Zhang2021} emphasize the 
open challenge: searching for critical executable scenarios from a very abstract one —a problem we try to address with \roadlogic.
Becker and Neurohr~\cite{becker2025correct} tried to address this problem by considering a simplified dynamic vehicle model (e.g., the bicycle model), vehicle trajectories as Bèzier splines, and encode abstract scenarios in a series of linear constraints systems that represent both temporal and spatial relations.  In contrast, \roadlogic uses ASP for the resolution of the logical constraints and in a high-level discrete-time plan where we can encode more easily specific driving rules.  Moreover, the work~\cite{becker2025correct} does support  
scenarios in OS2 like in \roadlogic and does not provide a monitor to check whether the simulation is compliant with the goal of the abstract scenario.

Wang et al.~\cite{WangXPJ23} also use ASP to generate collision-free ADS scenarios that satisfy logical constraints. In the paper, the authors mention that these ASP programs can be translated into OpenSCENARIO XML (1.0).   Similarly, Karimi et al.~\cite{karimi2022testcases} specify and solve complex logical constraints of autonomous driving scenarios, such as traffic rules, as ASP programs to automatically generate intersection test cases of increasing complexity, later executed in CARLA.  Klampfl and Wotawa demonstrate in~\cite{klampfl2024asp} the suitability of the ASP programs and reasoning for specifying high-level autonomous driving scenarios, proposing a novel approach to continuously monitor these specifications over data from the simulation environment to detect and explain potential discrepancies.  In contrast with these approaches, \roadlogic integrates ASP into a complete pipeline: takes input from an OS2
$\rightarrow$ symbolic automata $\rightarrow$ high-level planning via ASP $\rightarrow$ motion-planning-based concretization $\rightarrow$ specification monitoring in simulation.  This pipeline makes sure that generated scenarios remain both feasible and specification-compliant, bridging the gap between logical design and executable tests.

 Lin et al.~\cite{LinRA23} consider the problem of format interoperability between different scenario description languages by proposing a translator from OpenSCENARIO XML (1.x) to the CommonRoad format, enabling scenario simulation with CommonRoad tools. 
Interoperability is also key in \roadlogic, but within OS2.

Although some commercial tools support OS2 scenarios, such as Foretellix's Foretify toolchain\footnote{\url{https://www.foretellix.com/foretify-toolchain-overview/}}, ours is the first, to our knowledge, open-source toolchain framework that both supports OS2 and generates both a concrete scenario to simulate and a monitor in terms of symbolic automaton, verifying at runtime that the generated scenario is compliant with the original specification.

\section{Discussion}
\label{sec:discussion}

\begin{figure}[htb]
\includegraphics[width=\linewidth]{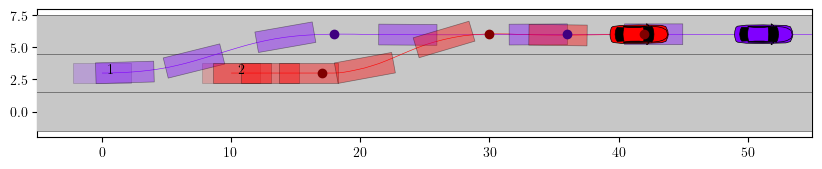}
\caption{A concrete behavior that satisfies the overtake scenario specification without being an actual overtake.}
\Description{}
\label{fig:unknown}
\end{figure}

The primary application of OS2 is large-scale verification and validation for assessing the safety and functionality of AVs. Its high level of abstraction and declarative style capture scenarios in terms of their intent, emphasizing what a scenario should achieve rather than how it is instantiated. This abstraction naturally supports systematic exploration of the scenario space to uncover potential unknowns. Therefore, generating concrete simulations from abstract scenarios provides a powerful mean for scenario elicitation.  Figure~\ref{fig:unknown} illustrates the detection of an unknown detected by \roadlogic from the overtake specification. We see that (1) the vehicle \ego is behind the vehicle \car in the middle lane at the beginning of the execution trace, (2) \ego is in the left lane with respect to \car in the middle of the trace, and (3) \ego is in the same (leftmost) lane as \car, but in front of it. It therefore satisfies the constraints of the overtake scenario. This behavior may however not meet the original intent of the scenario -- that \car does not change lane during the overtake maneuver. This additional constraint can be enforced in OS2, by restricting \car to drive straight, without changing the lane, as shown in Listing~\ref{lst:refined} (lines 7 and 8). 

\begin{lstlisting}[
    float=tp,
    floatplacement=tbp,
    basicstyle=\ttfamily\normalsize,
    language=OS2,
    caption={The overtake scenario refined with additional constraints preventing the vehicle \car from changing lane.},
    label={lst:refined}
]
scenario traffic.refined_overtake:
  v1: car
  v2: car

  do parallel():
    v2.drive() with:
      lane(0, at: start)
      lane(0, at, end)
    ...
\end{lstlisting}

The translation from a highly abstract and declarative formalism such as OS2 to concrete simulations also exposes subtle but impactful design choices. OS2 scenarios intentionally omit many low-level details such as vehicle dynamics or physical constraints that must nevertheless be resolved to obtain plausible and realistic simulations. In our approach, these missing aspects are addressed implicitly by representing domain knowledge in the form of additional ASP rules. This, however, raises an important question: what should remain explicit in the OS2 specification, and what should instead be treated as implicit, domain-wide knowledge applied uniformly across all scenarios?

\section{Conclusions and Future Work}
\label{sec:conclusions}

We introduced in this paper \roadlogic, the first open-source framework for automatically translating OS2 abstract autonomous driving scenario descriptions to concrete simulations. \roadlogic renders declarative driving scenarios executable, thus bridging the gap between the abstract and the concrete scenario layers.  

In this paper, we established the foundations for translating declarative OS2 specifications into concrete simulations, concentrating on the core maneuvers and the compositional operators used to construct complex scenarios. Adding many of the additional constructs from OS2 requires mainly engineering effort, which we plan to invest in improving our implementation. We also intend to support additional constructs, including events and non-highway road networks, which will require further research. 
While we demonstrated the effectiveness of our translation pipeline, we believe that the efficiency of the concrete scenario generation can be improved. We plan to optimize the ASP encoding of the scenarios and explore alternative planning methods, such as rapidly-exploring random trees (RRTs) to ensure more efficient simulation generation. Although we used specific simulation and motion planning environment to implement \roadlogic, our methodology does not depend on them. We hence intend to adapt \roadlogic to other AV environments such as Carla and BeamNG.

\paragraph{Acknowledgments.}
This research was funded in whole or in part by 
(i) the Austrian Science Fund (FWF) 10.55776/DOC1345324; 
(ii) the Vienna Science and Technology Fund (WWTF) under Grant ICT22-023 (TAIGER); 
(iii) the European Union, RobustifAI project, ID 101212818. Views and opinions expressed are however those of the author(s) only and do not necessarily reflect those of the European Union or the European Health and Digital Executive Agency (HADEA). Neither the European Union nor the granting authority can be held responsible for them;
(iv) the EU's Horizon Europe research and innovative programme under Grant Agreement No. 101160022 (VASSAL); and 
(v) the EU CHIPS JU Joint Undertaking under grant agreement n° 101194414 (project MOSAIC) and from the Austrian Research Promotion Agency (FFG).

\bibliographystyle{ACM-Reference-Format}
\bibliography{bibliography}

\end{document}